# Bound-state momentum-space wave function of the quasi-one-dimensional hydrogen atom

Aparna Saha[1], Benoy Talukdar[1] and Supriya Chatterjee[2]

[1] Department of Physics, Visva-Bharati University, Santiniketan 731235, India

[2] Department of Physics, Bidhannagar College, EB-2, Sector-1, Salt Lake, Kolkata 7000

**Abstract.** We prove that the bound-state momentum-space wave function $\varphi(p)$ for the quasi-one-dimensional hydrogen atom as used by Olendski in two recent publications (Eur. J. Phys.38, 038001(2017), arXiv : 1703.10042v1, 26 Mar2017) is incorrect. Further, we reconfirm that the wave function used by us in Eur. J.. Phys. 38, 025103 is correct.

The bound-state position-space wave function of the quasi-one-dimensional (Q1D) hydrogen atom is given by [1]

$$\psi(x) = \frac{2x}{\sqrt{n^3}} e^{-\frac{x}{n}} {}_1F_1\left(-n+1; 2; \frac{2x}{n}\right), \quad x \geq 0 \qquad (1)$$

with the energy eigenvalue written as

$$E_n = -\frac{1}{2n^2}. \qquad (2)$$

Here the principal quantum number $n = 1, 2, 3, etc$ and ${}_1F_1(.)$ stands for the confluent hypergeometric function. The wave function (1) can also be written in the equivalent form

$$\psi(x) = \frac{2x}{\sqrt{n^5}} e^{-\frac{x}{n}} L_{n-1}^{(1)}\left(\frac{2x}{n}\right) \qquad (3)$$

involving a generalized Laguerre polynomial $L_n^{(\beta)}(.)$ [2]. In close analogy with the result in (3) the eigenfunctions for the one-dimensional (1D) hydrogen atom $(-\infty < x < \infty)$ are given by [3]

$$\psi_e(x) = \sqrt{\frac{2}{n^5}} e^{-|x|/n} |x| L_{n-1}^{(1)}(2|x|/n) \qquad (4a)$$

and

$$\psi_o(x) = \sqrt{\frac{2}{n^5}} e^{-|x|/n} x L_{n-1}^{(1)}(2|x|/n), \qquad (4a))$$

where $\psi_e(x)$ stands for the wave function for the even states and $\psi_o(x)$, for the odd states. In a recent paper [4] we made use of (1) and a corresponding momentum- or $p$-space wave function written as

$$\varphi(p) = \frac{2^{\frac{5}{2}} n^{\frac{3}{2}}}{\sqrt{\pi}} \frac{p}{(1+n^2p^2)^2} U_{n-1}\left(\frac{1-n^2p^2}{1+n^2p^2}\right) \qquad (5)$$

with $U_m(.)$, the Chebyshev polynomial of the second kind [2] to compute results for position- and momentum-space Fisher information [5]. Our reason for assuming (5) as the correct momentum-space counterpart of (1) was the following.

The result in (1) can be obtained by using the angular momentum $l = 0$ in the radial part of the hydrogen-atom wave function. Thus it is quite plausible to write its momentum-space analog by putting $l = 0$ in the well known result of Podolsky and Pauling [6]. The coordinate-space wave function (1) vanishes at $x = 0$ and has additional $(n-1)$ zeros in the semi-infinite interval $0 \leq x \leq \infty$. As expected, the $p$-space wave function also exhibits a similar behavior [7] as a function of $p$.

In view of the above we felt that there is neither any physical uncertainty nor any mathematical flaw in working with the wave function in (5). Unfortunately, Olendski [8] could not accept (5) as a correct bound-state momentum-space wave function of Q1D hydrogen atom. In stead, he suggested that an alternative expression given by

$$\varphi^a(p) = (-1)^{n+1}\sqrt{\frac{2n}{\pi}}\frac{(1-inp)^{n-1}}{(1+inp)^{n+1}}, \quad -\infty < p < \infty \tag{6}$$

should be regarded as the right bound-state momentum-space wave function of the Q1D hydrogen atom. We have shown [9] that the imaginary part of $\varphi^a(p)$, namely,

$$\varphi_i^a(p) = (-1)^n\sqrt{\frac{2n}{\pi}}\frac{\sin(2n\arctan(np))}{1+n^2p^2} \tag{7}$$

for $p > 0$ is equal to $\varphi(p)$ in (5). Since $\varphi_i^a(p)$ is given in terms of elementary transcendental function, we recalculated the momentum-space Fisher information $I_\gamma$ by using the normalized form of the wave function (7) and verified that our result for $I_\gamma$ can be obtained from the corresponding result of the three-dimensional Coulomb problem [10] for $l=0$. Understandably, this provides a very useful check on our demand for (5) or (7) to represent the correct momentum-space wave function for the Q1D hydrogen atom. Despite that, Olendeski [11] have problems to accept our suggestion. While expressing his resentment Olehdski demonstrates that use of (7) in calculating Shannon's momentum-space information entropy leads to violation of entropic uncertainty relation [12].

The object of the present work is to explicitly demonstrate that $\varphi^a(p)$ is not the momentum-space wave function of Q1D hydrogen atom. We shall do this with particular attention to the position and momentum probability densities written in terms of the wave functions in (1) and (6). The same approach will be followed to establish that $\varphi^a(p)$ actually represents the correct $p$ - space wave function of the 1D hydrogen atom for which the representation-space wave functions are given in (4a) and (4b). Next we shall confirm that our suggested momentum-space wave function $\varphi(p)$ or $\varphi_i^a(p)$ stands for the correct result as demanded in refs. 4 and 9. In this context we shall make make appropriate comments on the violation of entropic uncertainty as noted in ref. 11.

In terms of wave functions in (1) and (6), the position- and momentum-space probability densities are given by

$$\rho(x) = \frac{4}{n^3}x^2 e^{-\frac{2x}{n}}{}_1F_1^2\left(-n+1;2;\frac{2x}{n}\right), \quad 0 \le x \le \infty \tag{8a}$$

and

$$\gamma(p) = \frac{2n}{\pi}\frac{1}{(1+n^2p^2)^2}, \quad -\infty \le p \le \infty . \tag{8b}$$

In Figure 1 we plot the position probability densities as a function of $x$ for $n=1$ and $2$. The solid curve gives the variation of $\rho_1(x)$ while the dashed curve gives similar variation of $\rho_2(x)$.

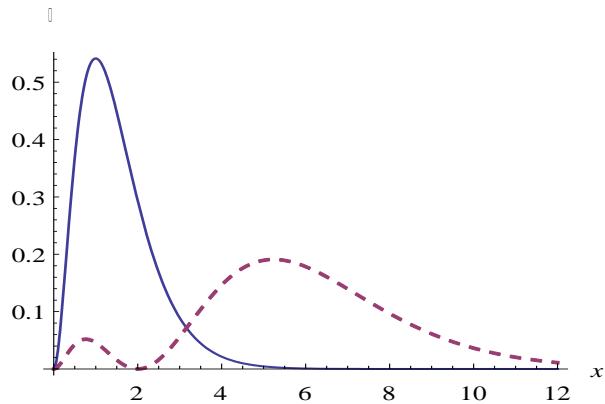

Figure 1. Position probability density $\rho(x)$ as a function of $x$, solid curve for $n=1$, dashed curve for $n=2$

As expected both $\rho_1(x)$ and $\rho_2(x)$ are zero at $x=0$, and $\rho_2(x)$ has an additional zero at $x \approx 2$. In Figure 2 we portray the momentum probability density $\gamma(p)$ from (8b) as a function of $p$. As before the solid and dashed curves give the variation of $\gamma_1(p)$ and $\gamma_2(p)$ respectively.

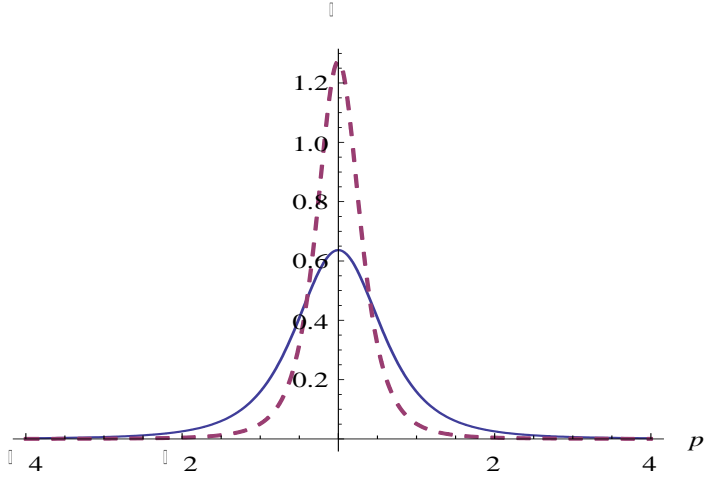

Figure 2. Momentum probability density $\gamma(p)$ as a function of $p$, solid curve for $n=1$, dashed curve for $n=2$

Both curves in this figure are Lorentzian, The dashed curve is more peaked compared to the solid one. This is understandable because the curve for $\rho_2(x)$ in Figure 1 is rather flat as compared to that for $\rho_1(x)$. In contrast to the curve for $\rho_1(x)$, the variation of $\rho_2(x)$ as a function of $x$ exhibits a structure, Ideally, one expects that the curve in the reciprocal space, namely the dashed curve in Figure 2 will bear some signature of the observed structure in Figure1, which has its origin at the node of $\psi(x)$ for $n=2$. Unfortunately, the probability density (8b) fails to do so. By dealing with still higher values on $n$ one can confirm that the probability density (8b) does not represent the correct momentum-space counterpart of (8a). In the following we demonstrate that (8b) actually stands for the momentum probability density of the 1D hydrogen atom.

From (4) we can write the position probability density for the 1D atom in the form

$$\rho(x) = \frac{2x^2}{n^5} e^{-2|x|/n} \left( L_{n-1}^{(1)}(2|x|/n) \right)^2, \quad -\infty \leq x \leq \infty. \tag{9}$$

In Figure 3 we present the plots of $\rho(x)$ in (9) for $n=1$ and $n=2$ as a function of $x$. The solid and dashed curves have the same meaning as used earlier. Both curves are zero at $x=0$. The solid curve exhibit two large humps, one lying at the positive quadrant and the other, at the negative quadrant. On the other hand, the dashed curve has two humps (one large and one small) in the positive quadrant and two other similar humps in the negative quadrant. The solid curve in Figure 2 giving the momentum probability distribution for $n=1$ is maximum at $p=0$ and vanishes on the wings. This curve thus truly represents the momentum counterpart of the ground-state position probability distribution of the 1D hydrogen atom shown in Figure 3. Arguments similar to those used for the ground-state probability distribution can be applied here to establish that the dashed curve in Figure 2 denotes the momentum probability distribution of the first exited state of the 1D hydrogen atom. In this context, we note that Davtyan et al [13] have shown the momentum-space wave function of the 1D hydrogen is given by

$$\varphi(p) = \sqrt{\frac{2n}{\pi}} \frac{e^{\pm 2in\arctan(np)}}{1+n^2p^2}. \tag{10}$$

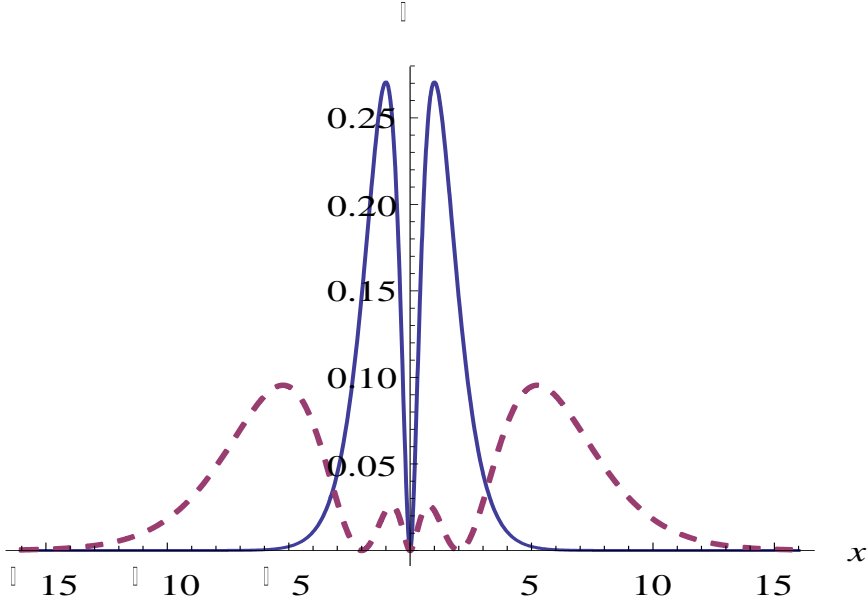

Figure 3, The position probability density for the 1D hydrogen atom as a function of $x$. The solid and dashed curves have the same meaning as used in Figure 2.

Clearly, the wave function (10) tells that (8b) is the momentum probability distribution of the 1D hydrogen atom and not of the Q1D atom. Moreover, we note that, to within a trivial phase factor, our proposed momentum-space wave function for the Q1D hydrogen atom in (5 ) or (7) is equal to the imaginary part of (10) for $p > 0$.

We shall now show that our proposed wave function (7) when normalized to unity yields the correct momentum probability distribution

$$\gamma(p) = \frac{8n}{\pi} \frac{\sin^2(2n\arctan(np))}{(1+n^2p^2)^2} \tag{11}$$

of the Q1D hydrogen atom. Figure 4 gives the plot of $\gamma(p)$ from (11) as a function of $p$ for $n=1$ and $n=2$. The curve of $\rho(x)$ for $n=1$ in Figure 1 is a highly peaked function. Consequently, the corresponding curve for the probability distribution in the reciprocal space should be relatively flat. Looking at the solid curve in Figure 4 we see that this is indeed the case. Comparing the dashed curves in Figures 1 and 4 we see that the structure of the curve for $\rho(x)$ is correctly reflected in the variation of $\gamma(p)$. For example, the small hump between lying $x=0$ and $x=2$ in the curve of $\rho(x)$ for $n=2$ in Figure 1 is converted to a huge hump in the reciprocal space. As is evident from the dashed curve in Figure 4, the maximum of the huge hump of $\gamma(p)$ is situated at $p \approx 0.18$. Moreover, the dashed curve touches the $p$-axis at $p=0.5$ presumably because the $2s$ wave function of the Q1D hydrogen atom has a node there. The small hump in the outer region of the dashed curve is associated with the huge hump in the curve for $n=2$ in Figure 1. This analysis firmly establishes that the expression in (5) or (7) represents the correct momentum-space wave function of the Q1D hydrogen atom. Some comments on the violation of the entropic uncertainty relation [13] as noted by Olendski [11] are now in order.

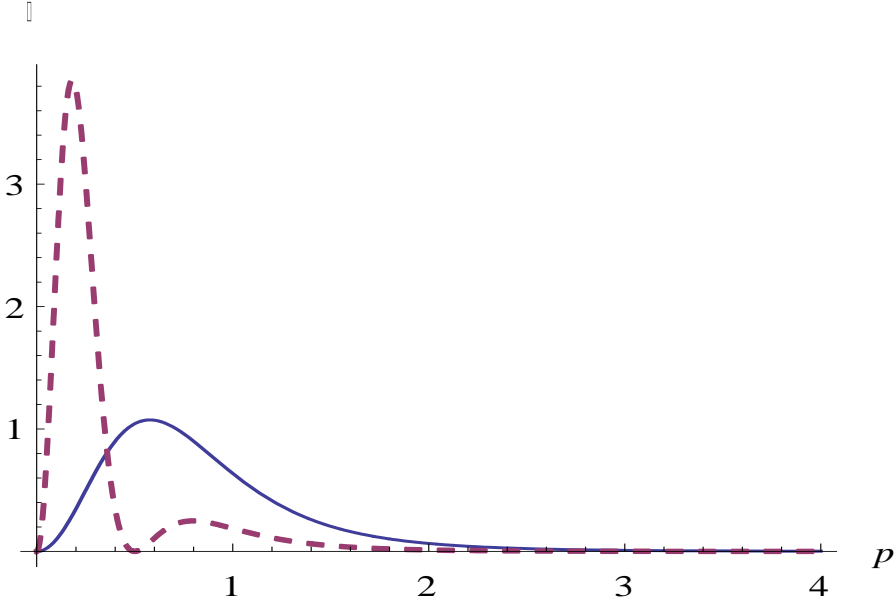

Figure 4. Momentum probability distribution for the Q1D hydrogen for n=1 and 2 as a function of p using our momentum-space wave function (7). The solid and the dashed curve as used in earlier figures.

For 1D system the entropic uncertainty relation or the so-called BBM inequality [12] is given by

$$S_\rho + S_\gamma \geq 1 + \ln \pi,  \qquad (12)$$

where the position- and momentum-space information entropies are defined as [14]

$$S_\rho = -\int_0^\infty \rho(x) \ln \rho(x) dx \qquad (13a)$$

and

$$S_\gamma = -\int_a^b \gamma(p) \ln \gamma(p) dp. \qquad (13b)$$

There is no problem to recognize that the position probability in (13a) for the Q1D atom is given by (9) with $0 \leq x \leq \infty$. But we have seen that so far (13b) is concerned there exists two conflicting view points regarding the choice of $\gamma(p)$. (i) Olendski [8,11] believes that the correct expression for the momentum probability density is given by (8b). (ii) Contrarily, we have provided a number of evidences to confirm that $\gamma(p)$ in (11) represents the correct momentum probability density of the Q1D hydrogen atom. For (i) the limits of integration in (13b) are given by $a = -\infty$ and $b = +\infty$. Here the ortho-normality relation of the bound-state position- and momentum-space eigenfunctions is given by

$$\int_0^\infty \psi_n(x)\psi_{n'}(x)dx = \int_{-\infty}^\infty \varphi_n^*(p)\varphi_{n'}(p)dp = \delta_{nn'}. \qquad (14)$$

Equation (14) violates the well known Perseval theorem because the right side of it involves eigenfunctions of the Q1D system while in middle we have eigenfunctions of the 1D system. For our chosen of $\gamma(p)$ in (11) $a=0$ and $b=\infty$ such that in this case the ortho-normality relation is consistent with the Perseval theorem. From (8a), the value of $S_\rho$ for $n=1$ can be obtained as 1.1511. This result is same for both (i) and (ii). According to Olendski we can make use (8b) with $n=1$ to obtain $S_\gamma = 1.2242$ from (13b) such that the entropy sum is given by

$$S_\rho + S_\gamma = 2.3786 . \qquad (15)$$

Apparently, this is a very satisfying result because the sum in (15) $> 1 + \ln \pi$ and thus does not violate the entropic uncertainty relation. But if we note that in the above we used the value of $S_\rho$ for the Q1D atom and that of $S_\gamma$ for the 1D atom, the claim regarding the validity of uncertainty becomes meaningless. On the other hand, if we work with the true momentum probability density of the Q1D atom we get from (13 b) the value of $S_\gamma$ as $S_\gamma = 0.5575$ leading to $S_\rho + S_\gamma = 1.7119$. This result is less than $1 + \ln \pi$ $(= 2.1447)$ Thus we find here an obvious violation of entropic uncertainty. Since there exists a vast amount of literature [15] on the applications of entropic uncertainty relation of quantum mechanical systems as an alternative to Heisenberg uncertainty principle, the violation of the relation in (12) noted here needs some serious consideration.

In close analogy with the entropic uncertainty relation (12), for 1D systems there exists a Fisher-information based uncertainty relation given by [16].

\  $$I_\rho I_\gamma \geq 4, \qquad (16)$$

where $I_\rho$ and $I_\gamma$ stand for the position- and momentum-space Fisher information [17]. In general, the result in (16) has been found to be valid for time-independent systems. However, as opposed to the conjecture of Hall [18], the results of Fisher information for time-dependent systems [19] were used to conclude that one cannot write an universal uncertainty relation as a lower bound to the product $I_\rho I_\gamma$. In addition to time-dependent systems, Agular and Guedes [20] found that Fisher information of damped harmonic oscillator violates the uncertainty relation in (16). But there exist no such physical examples which invalidate the entropic uncertainty relation (12). However, one can cite many elementary examples for which the BBM inequality gets saturated. One such example is provided by the ground state of the harmonic oscillator [21]. Also the inequality for the entropic uncertainty relation gets exhausted for the ground states of Morse and Poschl-Teller potentials [22].

In the above we noted that for $n=1$ the momentum density (11) gives a value for $I_\gamma$ which in conjunction with the corresponding position-space result leads to violation of the entropic uncertainty . Since (11) represents the correct expression for $I_\gamma$ of the Q1D hydrogen atom, we can make judicious use of the terminology in refs. 21 and 22 to go one step forward and say that the sum $I_\rho + I_\gamma$ for the ground-state of the Q1D atom is supersaturated. It will be of some interest to examine how does the entropy sum behave as one moves to higher excited states. In view of this we present in Table 1 the results for Shannon information entropies of the Q1D hydrogen atom for $n=1$ to $10$. Equation (10) confirms that the expression for $\gamma(p)$ in (8b) used by Olendski [11] to compute the result for $S_\gamma$ actually stands for the momentum probability distribution of the 1D hydrogen atom. Despite that we have presented in column 3 of Table 1 the values of momentum-space information entropies computed by following the viewpoint of ref. [11] and denoted them by $S_\gamma^o$. Column 5 of the table gives the entopic sum $S_\rho + S_\gamma^o$ as a function of $n$. So far as BBM inequality is concerned the results for the entropy sum appear to be quite satisfactory. But it is clear that these numbers do not carry any physical meaning since the values of $S_\rho$ are computed for the Q1D atom and those of $S_\gamma^o$, for the 1D atom. Since in ref. 8 the momentum-space Fisher information was calculated using the same momentum probability distribution, (17b) of ref.8 is also incorrect. The values in column 4 represent the actual results of the momentum-space entropies ($S_\gamma^s$) of the Q1D hydrogen atom. These results have been calculated by taking recourse to the use of the expression for $\gamma(p)$ in (11). The corresponding entropy sum $S_\rho + S_\gamma^s$ is shown in column 6 of the table. From the numbers presented we see that the values of the sum increases almost monotonically with $n$. Clearly, the entropic uncertainty is violated for $n=1$ and $2$. The entropy sum tends to saturate at $n=3$. Henceforth all results satisfy the BBM inequality. It remains an interesting curiosity to investigate these points in some detail and thereby look into the physical origin for the violation of entropic uncertainty. However, as regards the correct form of the momentum-space wave function and/ or the corresponding probability density we conclude by noting the following.

Table 1. Shannon information entropies of the Q1D hydrogen atom for principal quantum numbers n=1 to10

| n | $S_\rho$ | $S_\gamma^o$ | $S_\gamma^s$ | $S_\rho + S_\gamma^o$ | $S_\rho + S_\gamma^s$ |
|---|---|---|---|---|---|
| 1 | 1.1544 | 1.2242 | 0.5575 | 2.2786 | 1.7119 |
| 2 | 2.2343 | 0.5310 | -0.4357 | 2.7653 | 1.7968 |
| 3 | 2.9056 | 0.1256 | -0.8668 | 3.0312 | 2.0388 |
| 4 | 3.3954 | -0,1621 | -1.1622 | 3.2333 | 2.2332 |
| 5 | 3.7817 | -0.3853 | -1.3947 | 3.3964 | 2.3870 |
| 6 | 4.1012 | -0.5676 | -1,5900 | 3.5336 | 2.5112 |
| 7 | 4.3737 | -0.7217 | -1.7407 | 3.6520 | 2.6333 |
| 8 | 4.6114 | -0.8553 | -1.8760 | 3.7561 | 2.7354 |
| 9 | 4.8223 | -0.9830 | -2.0065 | 3.8393 | 2.8158 |
| 10 | 5.0118 | -1.0784 | -2.1055 | 3.9334 | 2.9063 |

About two decades ago Vos and McCarthy [23] in a very interesting paper on electron-momentum spectroscopy pointed out, 'There are a lot of similarities between the orbitals in momentum and coordinate space. The symmetry properties of each orbital are identical in both representations; even the shapes are similar.' For hydrogenic wave function these similarities can easily be verified using the results given in ref. 7. For the Q1D hydrogen atom we had chosen to work with similar coordinate- and momentum-space wave functions [4]. Olendski [8] did not agree with our choice and thus suggested an entirely different form for $\varphi(p)$. In this work we proved that as a consequence of (10), $\varphi^a(p)$ in (6) does not represent a physically acceptable wave function of the Q1D hydrogen atom. Accidently, its absolute square gives the momentum probability distribution for the 1D hydrogen atom.